\documentclass{llncs}

\usepackage[utf8]{inputenc}
\usepackage{amsmath}
\usepackage{amsfonts}
\usepackage{amssymb}
\usepackage{graphicx}
\usepackage[ruled,vlined,linesnumbered]{algorithm2e}

\def \calC {\mathcal{C}}
\def \calE {\mathcal{E}}
\def \calF {\mathcal{F}}
\def \calG {\mathcal{G}}
\def \calN {\mathcal{N}}
\def \calP {\mathcal{P}}

\def \calR {\mathcal{R}}

\title{Evolution of Communities with Focus on Stability}

\author{Carlos Sarraute \and Gervasio Calderon}
\institute{Grandata Labs \\
Buenos Aires, Argentina \\
\email{ \{charles,gerva\}@grandata.com}
}
\date{}

\begin{document}
\maketitle

\begin{abstract}

Community detection is an important tool for analyzing the social graph of
mobile phone users.
The problem of finding communities in static graphs has been widely studied.
However, since mobile social networks evolve over time, static graph algorithms
are not sufficient.
To be useful in practice (e.g. when used by a telecom analyst),
the stability of the partitions becomes critical.

We tackle this particular use case in this paper:
tracking evolution of communities in dynamic scenarios with focus on stability.
We propose two modifications to a widely used static community detection
algorithm: we introduce fixed nodes and preferential attachment to pre-existing 
communities.
We then describe experiments to study the stability and quality of the
resulting partitions on real-world social networks, represented by monthly
call graphs for millions of subscribers.

\end{abstract}

\section{Introduction}

In recent years, there has been a growing interest in
the scientific community and mobile phone operators to
analyze the huge datasets generated by these operators,
for scientific and commercial purposes (e.g. as demonstrated by the growth of
NetMob attendance).

One of the important tools used to analyze the social graph of mobile phone
users is community detection.
Within each community, external nodes are susceptible of attraction to become
customers and
existing customers can be influenced to remain and/or to buy new products
or services.
This can be done by leveraging mutual influence of nodes \cite{wu2009group}.
User communities, as smaller units easier to manage, allow the computation of role
analysis. The centrality of an actor within a community is the base for role
analysis.

The problem of finding communities in static graphs has been widely studied
(see \cite{fortunato2010community} for a survey).
However, for practical use cases, the detected communities must evolve matching
the underlying social graph of communications (for example, to track ongoing marketing
campaigns aimed at specific communities).
Also, behavior analysis of communities of users over time
can be used to predict future activity that interests telecom operators,
such as subscriber churn or handset adoption \cite{greene2010tracking}.
Similarly, group evolution can provide insights for designing strategies
such as early warning of group churn.

The stability of communities is critical to preserve earlier analysis.
This is the specific use case we tackle in this paper:
tracking evolution of communities in dynamic scenarios with focus on stability.

The rest of the paper is organized as follows. In Section
\ref{sec:data_sources}
we describe the data sources we used to perform our experiments.
In Section \ref{sec:dynamic} we propose two modifications to a widely used
static community detection algorithm (the Louvain Method of Vincent Blondel
and his team \cite{blondel2008fast}).
Then, in Section \ref{sec:results}, we describe the experiments performed in
order to study the stability and quality of the resulting partitions.
These experiments were run on real-world social networks represented by
monthly call graphs for millions of subscribers.
Section \ref{sec:related-work} reviews related work, and Section \ref{sec:conclusion} 
concludes the paper with ideas for future work.

\section{Data Sources} \label{sec:data_sources}

Our raw data source is anonymized traffic information from a mobile operator.
The analyzed information ranges from January 2012 to January 2013 (same time span as \cite{wu2009group}).
For each communication it contains the origin, target, date and time of
the call or sms, and duration in the case of calls.

For each month $T$, we construct a social graph
$\calG_{T} = < \calN_{T}, \calE_{T} > $.
This graph is based on the aggregation of traffic for several months,
more concretely $\calG_{T}$ depends on the traffic of three months:
$T$, $T-1$ and $T-2$.
The raw aggregation of calls and messages gives a first graph
with around 92~M (million) nodes and 565~M edges (on a typical month).
Voice communications contribute 413~M edges and text messages contribute
296~M edges to this graph.

Then, we perform a symmetrization of the graph keeping only the edges $(A,B)$
when there are communications from $A$ to $B$ and from $B$ to $A$.
This new graph contains around 56~M nodes and 133~M (undirected) edges,
representing stronger social interactions between nodes.
Additionally we filter nodes with high degree (i.e. degree greater than 200)
since we are interested in the communications between people (and not call
centers or platform numbers).

\section{Dynamic Louvain Method} \label{sec:dynamic}

Our first experiment to detect evolving communities was to run the 
original Louvain algorithm \cite{blondel2008fast} on the 
graphs at time $T$ and $T+1$ and compare the two partitions. This
method produced very unstable results.
Our second experiment was to run the Louvain algorithm with the modifications
by Aynaud and Guillaume \cite{aynaud2010static} (denoted LMAG hereafter)
to obtain a more stable evolution.

This is our implementation of the initial step of LMAG. At time $T+1$, the
nodes already present at time $T$ are initially assigned to the community
they belonged to at time $T$. New nodes -- not present at time $T$ -- are
assigned to fresh communities (see Algorithm \ref{algo-LMAG}).
As we show in Section~\ref{sec:results}, results were still
unsuitable in terms of stability.

\begin{figure*}[t]
\centering
\begin{algorithm}[H]
\SetKwData{InitialComms}{initialComms}
\SetKwData{MaxOldCommunity}{maxOldCommunity}
\SetKwData{AvailableComm}{availableComm}
\SetKwFunction{OldComm}{oldComm}
\SetKwFunction{Max}{Max}
\SetKwFunction{IsRandomFreeNode}{IsRandomFreeNode}
\SetKwFunction{FindInOldCommunities}{FindInOldCommunities}
\KwIn{$r$ (percentage of free nodes). $\calN_{T+1}$ (nodes). $\calN_{T} \rightarrow \cal$C$_{T}$ (old nodes to communities)}
\KwOut{Initial communities assigned to nodes}
\MaxOldCommunity$\leftarrow$ \Max{$\cal$C$_{T}$}\;
\AvailableComm$\leftarrow$ \MaxOldCommunity + 1\;
\ForEach{node x in $\calN_{T+1}$}{
	\OldComm$\leftarrow$ \FindInOldCommunities{$x$}\;
	\eIf{\OldComm is not NULL and not \IsRandomFreeNode{$r$} }{
		\InitialComms[$x$] $\leftarrow$ \OldComm\;
	}{
		\InitialComms[$x$] $\leftarrow$ \AvailableComm\;
		\AvailableComm $\leftarrow$ \AvailableComm + 1\;
	}
}
\Return \InitialComms
\caption{Initial step of Louvain modified by Aynaud and Guillaume}
\label{algo-LMAG}
\end{algorithm}
\end{figure*}

In our use case (e.g. telecom analysts performing actions on communities),
the stability of the partition is our main concern (see stable or natural communities in \cite{fortunato2010community}).
With this goal in mind, we propose two modifications to the Louvain method,
that give the partition at the previous time step a sort of ``momentum",
and make it more suitable to track communities in dynamic graphs.

Before describing them, we introduce some notations.
As stated in the previous section, we consider snapshots of the social graph
constructed at discrete time steps (in our case every month).
Let $\calG_{T} = < \calN_{T}, \calE_{T} >$ be a graph that has already
been analyzed and partitioned in communities. Let $\Gamma = <C_1, \ldots, C_R>$
be such partition in $R$ communities.
Given a new graph $\calG_{T+1} = < \calN_{T+1}, \calE_{T+1} >$
our objective is to find a partition of $\calG_{T+1}$ which
is stable respect to $\Gamma$.

The first idea is to have a set of \emph{fixed nodes} $\calF$.
Let $\calR = \calN_{T} \cap \calN_{T+1}$  be the set of nodes that remain from 
time $T$ to $T+1$.
The set $\calF$ is a subset of $\calR$, whose  
nodes are assigned to the community they had at time $T$.
In other words, noting $\gamma$ the function that assigns a community to each node, 
we require: 
$ \gamma_{T+1}(x) = \gamma_{T}(x) \; \forall x \in \calF $.

We experimented with different distributions of the fixed nodes,
ranging from no fixed nodes ($\calF = \emptyset$) to all the remaining nodes 
($\calF = \calR$). 
For the experimental results, we used a parameter $p$
representing the probability that a node belongs to $\calF$
(i.e. $| \calF | = p \cdot | \calR | $).

The second idea is to add a probability $q$ of ``preferential attachment"
to pre-existing communities.
With probability $q$, new nodes will be more likely to attach to an existing
community at time $T$ instead of attaching to a community formed at time $T+1$.
We give the details below.

The Louvain Method \cite{blondel2008fast} is a hierarchical greedy algorithm
composed of two phases.
During phase 1, nodes are considered one by one and each one is placed
in the neighboring community (including its own community)
that maximizes the modularity gain.
This phase is repeated until no node is moved (that is when the decomposition
reaches a local maximum).
Phase 2 consists of building the graph between the communities obtained
during phase 1.
Then, the algorithm starts phase 1 again with the new graph, in the next
hierarchical level of execution, and continues until the modularity does not
improve anymore.

We construct a set $\calP \subseteq \calN_{T+1}$
such that $|\calP| = q \cdot | \calN_{T+1} |$.
For every node $x$, we consider its neighbors that belong to a community existing 
at time $T$, that is the set 
$ A(x) = \{ z \in \calN_{T+1} \, | \, (x,z) \in \calE_{T+1} \wedge \gamma_{T+1}(z) \in \Gamma_T  \} $.
During phase 1 of the first iteration of the algorithm (i.e. during the first
hierarchical level of execution), the inner loop is modified.
For all node $x \in \calN_{T+1}$, if $x \in \calP$ and $A(x) \neq \emptyset$ then place $x$ in the community
of $A(x)$ which maximizes the modularity gain
(whereas if $A(x) = \emptyset$ proceed as usual).

The two methods (``fixed nodes" and ``preferential attachment") apply at every Louvain step and, since they are related, may be described as a whole (see Algorithm \ref{algo-dynamic}).

\begin{figure*}[t]
\centering
\begin{algorithm}[H]
\SetKwData{OldNodesToComms}{oldNodesToComms}
\SetKwData{NodesToComms}{nodesToComms}
\SetKwData{AvailableComm}{availableComm}
\SetKwData{X}{x}
\SetKwData{OldComm}{oldComm}
\SetKwData{OldNeighborComm}{oldNeighborComm}
\SetKwData{ModularityIncrease}{modularityIncrease}
\SetKwData{BestOldIncrease}{bestOldIncrease}
\SetKwData{BestNewIncrease}{bestNewIncrease}
\SetKwData{BestIncrease}{bestIncrease}
\SetKwFunction{Max}{Max}
\SetKwFunction{IsRandomFixedNode}{IsRandomFixedNode}
\SetKwFunction{FindInOldCommunities}{FindInOldCommunities}
\KwIn{Probabilities p, q. $\calN_{T} \rightarrow \calC_{T}$ (old nodes to communities). Nodes $\calN_{T+1}$. }
\KwOut{Communities assigned to nodes at this Louvain step.}
\ForEach{node x in $\calN_{T+1}$}{
\OldComm $\leftarrow$ \FindInOldCommunities{$x$}\;
\tcp{FIXED NODES (p)}
\eIf{\OldComm is not NULL and \IsRandomFixedNode{$p$}}{
Do nothing. Keep the old community\;
}{
\ForEach{neighbor $n$ of $x$}{
\ModularityIncrease $\leftarrow$ Calculate increase in modularity were x to be moved to n's community\;
\eIf{$n$ is in old communities}{
\BestOldIncrease $\leftarrow$ \ModularityIncrease if bigger;
}{
\BestNewIncrease $\leftarrow$ \ModularityIncrease if bigger;
}
}
\tcp{PREFERENTIAL ATTACHMENT (q)}
\eIf{$x$ has preferential attachment to old communities (using $q$ factor)}{
\BestIncrease $\leftarrow$ \BestOldIncrease if there's at least one ``old" neighbor else \BestNewIncrease;
}{
\BestIncrease $\leftarrow$ \Max{\BestOldIncrease, \BestNewIncrease};
}
Move $x$ from current community to \BestIncrease one, if different.
}
}
\Return \NodesToComms
\caption{Fixed nodes and Preferential attachment}
\label{algo-dynamic}
\end{algorithm}
\end{figure*}

\section{Experimental Results} \label{sec:results}

\begin{figure}[ht]
\centerline{\includegraphics[width=0.80 \textwidth]{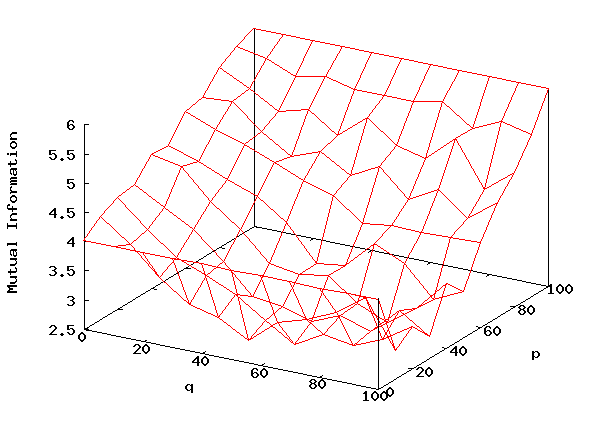}}
  \caption{Mutual Information  as a function of $p$ and $q$ (expressed as percentages).}
  \label{fig:mutual_information}
\end{figure}

\begin{figure}[ht]
\centerline{\includegraphics[width=0.80 \textwidth]{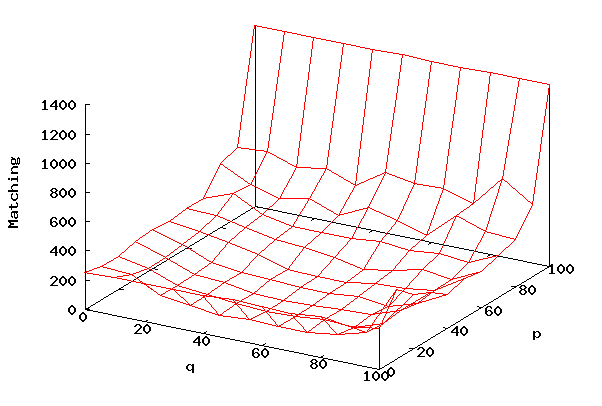}}
  \caption{Matching communities as a function of $p$ and $q$.}
  \label{fig:matching}
\end{figure}

\begin{figure}[ht]
\centerline{\includegraphics[width=0.80 \textwidth]{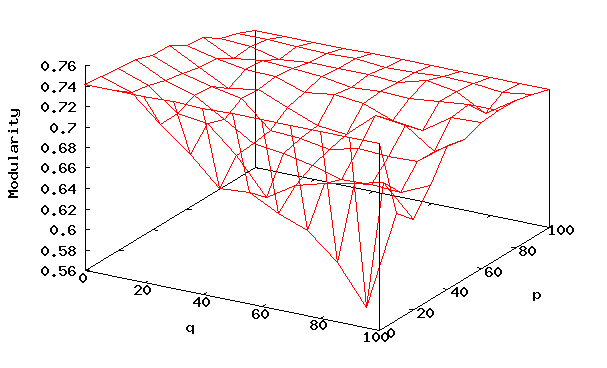}}
  \caption{Modularity as a function of $p$ and $q$.}
  \label{fig:modularity}
\end{figure}

In our experiments we computed the social graph 
(constructed as described in Section \ref{sec:data_sources}).
Since we are interested in the real-world application of our method
we preferred to evaluate it on real data.

Given two months $T$ and $T+1$, we calculated a partition in communities
of $\calG_{T}$ using the Louvain Method (with the modification of \cite{aynaud2010static})
that we note $\Gamma = < C_1, \ldots, C_R > $;
and a partition of $\calG_{T+1}$ using our dynamic version of the Louvain Method,
with different values of the parameters $p$ and $q$. 
Let $\Gamma' = < C'_1, \ldots, C'_S > $ be the partition of $\calG_{T+1}$.
We are interested in comparing $\Gamma$ and $\Gamma'$ in terms of stability
and quality of the partition. To this end, we measure:
 (i) the mutual information between the two partitions;
 (ii) the number of matching communities (i.e. such that the proportion 
 of nodes in common is greater than a parameter $r$);
 (iii) the final modularity of $\Gamma'$ (as defined in \cite{newman2004finding}).

The number of matching communities is computed as follows: for each
community $C'_j \in \Gamma'$, we evaluate whether there is a community
$C_i \in \Gamma$ such that $ | C_i \cap C'_j | > r \cdot | C_i | $
and $ | C_i \cap C'_j | > r \cdot | C'_j | $, where $r$ is a fixed
parameter verifying $r > 0.50$ (for instance we used $r = 0.51$).
In that case, we say that $C'_j$ matches $C_i$ (another criterion is Jaccard similarity:
intersection size divided by union size, see \cite{fortunato2010community} and \cite{wu2009group}).
The matching communities are of particular interest,
because $C'_j$ can be considered as the evolution of $C_i$ (although the community
may have grown or shrank) and can be individually followed by a human analyst.

The mutual information for two partitions of communities
(see \cite{fenn2012dynamical,greene2010tracking} for definitions\footnote{Since
nodes can change between time $T$ and $T+1$, we only consider the intersection
$\calN_{T} \cap \calN_{T+1}$ for the mutual information computation.}) 
is computed as: 
$$
MI(\Gamma,\Gamma') = \sum_{i=1}^{R} \sum_{j=1}^{S} P(C_i,C'_j) 
\log \frac{ P(C_i,C'_j) }{ P(C_i) \cdot P(C'_j) } \enspace .
$$

To analyze the effect of $p$ and $q$, we made those parameters vary from $0$ to $1$.
The baseline, for $p=0$ and $q=0$, corresponds to the Louvain Method
with the modifications of \cite{aynaud2010static}.

Fig.~\ref{fig:mutual_information} shows the effect on the mutual information
between the two partitions. We can clearly observe that the mutual information
increases as $p$ increases and reaches its maximal values at $p = 100\%$.
The effect of varying $q$ is not so clear since it produces fluctuations
of the mutual information without a marked tendency.

Fig.~\ref{fig:matching} shows the number of matching communities (according to our
criterion). In this graph we see that the number of matching communities 
increases dramatically when $p$ approaches $100\%$.
The effect of varying $q$ is again not clearly marked although the increase
of $q$ produces higher matching communities for smaller values of $p$.

Fig.~\ref{fig:modularity} shows the effect on the modularity of the new partition.
We can observe that the modularity decreases slightly as $p$ increases
for small values of $q$. For greater values of $q$ (closer to $100\%$),
varying $p$ produces fluctuations with a decreasing tendency.

As a conclusion, we can see that increasing the probability $p$ of fixed nodes
has a clear effect on increasing the mutual information between the two partitions
and the number of matching communities. The trade-off with quality is good,
since the decrease in modularity is relatively low.

By contrast, increasing the probability $q$ of preferential attachment
to pre-existing communities has no clear effect on mutual information or matching communities.
It does not seem advisable to use this second modification to generate
evolving communities.

\section{Related Work} \label{sec:related-work}

Many related papers have investigated the importance of communities, and served us as 
reference.
Foundational concepts and examples for communities detection may be found in \cite{newman2004finding}. For instance, the interactions between the major characters 
of the novel ``Les Misérables'' by Victor Hugo can be viewed as a graph with 77 nodes, 
and the characters organized as communities (the example is
very intuitive and easy to grasp for the fans of the novel). 
Another classic example is Zachary’s karate club, an organization with 34 nodes that
split into two separate clubs in real life. 
Each new part can be mapped almost perfectly to the 
two main communities of friends detected in the original club.
The paper also analyzes algorithms, such as shortest-path and random walk. 
Of course, the graphs considered in \cite{newman2004finding} have a much smaller 
scale than the ones considered in this work.

To perform community detection in graphs with 92 million nodes (see Section~\ref{sec:data_sources}), efficient algorithms are required.
We based our research on the Louvain Method algorithm
originally published in \cite{blondel2008fast}.
As discussed in Section~\ref{sec:dynamic}, it was modified in \cite{aynaud2010static}
to get a dynamic algorithm. However, that algorithm still lacks stability.
We used the implementation of \cite{aynaud2010static} as baseline to
evaluate our Dynamic Louvain Method.

A thorough study of the history and the state of the art in communities detection (up to 2010) can be found in \cite{fortunato2010community}.
In particular the author discusses ideas on the roles of vertices within communities.
We have implemented a classification of nodes as leaders, followers and marginals within
each community. Our leaders correspond to ``central vertices", 
but we don't compute boundary vertices, which could be useful.

Besides the static analysis, the report discusses communities evolution (dynamic communities), although it points out that
``the analysis of dynamic communities is still in its infancy." 
It suggest that it would be desirable ``to have a
unified framework, in which clusters are deduced both from the current structure of 
the graph and from the knowledge of the cluster structure at previous times." 
We have implemented that idea, since we use the previous history of community structure
throughout the whole algorithm (according to the $p$ and $q$ parameters), 
and not only during the nodes initialization (as in \cite{aynaud2010static}).

Applications of ``communities evolution" are not only to be found in mobile social networks.
In \cite{palla2007quantifying}, the authors study the evolution of 
scientific collaboration networks.
In the analysis of exchange markets \cite{fenn2012dynamical}, the dynamics of 
currency exchanges (viewed as a dynamic graph of currency pairs) have been studied. 
For instance, modifications in the currency exchange communities effectively reflect 
the Mexican peso crisis of 1994. 
The scale is also much smaller (only 11 currencies being analyzed). 
A similarity with our work is that financial markets are one of the few fields 
where a detailed time evolution is readily available. In our case we have data from 
telecom companies that span a wide range of time (several months), 
and has fine grained resolution (day, hour, minute and second of each call or message).

\section{Conclusion and Future Steps} \label{sec:conclusion}

The  detection of evolving communities is a subject that still requires further
study from the scientific community.
We propose here a practical approach for a particular version of this problem
where the focus is on stability.
The introduction of fixed nodes (with probability $p$)
increases significantly the stability of successive partitions,
at the cost of a slight decrease in the final modularity of each partition.

As future steps of this research, we plan to:
(i) study the evolution of communities 
with finer grain, using smaller time steps;
(ii) evaluate the proposed method on publicly available datasets,
to facilitate the comparison of our results;
(iii) refine the matching criteria, and consider additional events
in the evolution of dynamic communities (such as 
birth, death, merging, splitting, expansion and contraction \cite{greene2010tracking}).


\bibliographystyle{unsrt}

\bibliography{sna}

\begin{thebibliography}{1}

\bibitem{wu2009group}
Bin Wu, Qi~Ye, Shengqi Yang, and Bai Wang.
\newblock Group {CRM}: a new telecom {CRM} framework from social network
  perspective.
\newblock In {\em CNIKM'09}, pages 3--10. ACM, 2009.

\bibitem{fortunato2010community}
Santo Fortunato.
\newblock Community detection in graphs.
\newblock {\em Physics Reports}, 486(3):75--174, 2010.

\bibitem{greene2010tracking}
Derek Greene, D{\'o}nal Doyle, and P{\'a}draig Cunningham.
\newblock Tracking the evolution of communities in dynamic social networks.
\newblock In {\em ASONAM'10}, pages 176--183. IEEE, 2010.

\bibitem{blondel2008fast}
Vincent Blondel, Jean-Loup Guillaume, Renaud Lambiotte, and Etienne Lefebvre.
\newblock Fast unfolding of communities in large networks.
\newblock {\em Journal of Statistical Mechanics: Theory and Experiment},
  2008(10):P10008, 2008.

\bibitem{aynaud2010static}
Thomas Aynaud and Jean-Loup Guillaume.
\newblock Static community detection algorithms for evolving networks.
\newblock In {\em WiOpt'10}, pages 513--519. IEEE, 2010.

\bibitem{newman2004finding}
Mark~EJ Newman and Michelle Girvan.
\newblock Finding and evaluating community structure in networks.
\newblock {\em Physical review E}, 69(2):026113, 2004.

\bibitem{fenn2012dynamical}
Daniel~J Fenn, Mason~A Porter, Peter~J Mucha, Mark McDonald, Stacy Williams,
  Neil~F Johnson, and Nick~S Jones.
\newblock Dynamical clustering of exchange rates.
\newblock {\em Exchange Organizational Behavior Teaching Journal}, 2012.

\bibitem{palla2007quantifying}
Gergely Palla, Albert-L{\'a}szl{\'o} Barab{\'a}si, and Tam{\'a}s Vicsek.
\newblock Quantifying social group evolution.
\newblock {\em Nature}, 446(7136):664--667, 2007.

\end{thebibliography}

\end{document}